\documentstyle[twocolumn,aps,psfig,epsfig]{revtex}
\begin{document} 
\draft
\title{Electromagnetic induction and damping \\
-- quantitative experiments using PC interface}
\author{Avinash Singh, Y. N. Mohapatra, and Satyendra Kumar}
\address{Department of Physics, Indian Institute of Technology Kanpur - 208016, India}
\maketitle
\begin{abstract} 
A bar magnet, attached to an oscillating system,
passes through a coil periodically, generating a series of emf pulses.
A novel method is described for the 
quantitative verification of Faraday's law
which eliminates all errors associated with angular measurements,
thereby revealing delicate features of the underlying mechanics.
When electromagnetic damping is activated by short-circuiting the coil,
a distinctly linear decay of oscillation amplitude is surprisingly observed.
A quantitative analysis reveals an interesting interplay of the
electromagnetic and mechanical time scales. 

\end{abstract}
\section{Introduction}
Laboratory experiments on Faraday's law of electromagnetic induction
most often involve a bar magnet moving (falling) through a coil, 
and studying the induced emf pulse.\cite{emf1,emf2,emf3,emf4}
Several parameters can be varied, such as the velocity of the magnet, 
the number of turns in the coil, and the strength of the bar magnet.
The observed proportionality of the peak induced emf
on the number of turns in the coil and the magnet velocity provide
a quantitative verification of the Faraday's law. 

Commonly it is found convenient to attach the magnet to an
oscillating system, so that it passes through the coil
periodically, generating a series of emf pulses.
This allows the peak emf to be easily determined 
by charging a capacitor with the rectified coil output.
A simple, yet remarkably robust setup which utilizes this concept 
involves a rigid semi-circular frame of aluminum, 
pivoted at the center (O) of the circle (see Fig. 1). 
The whole frame can oscillate freely in its own plane about a
horizontal axis passing through O.
A rectangular bar magnet is mounted at the center of the arc and the arc passes
through a coil C of suitable area of 
cross section.\footnote{In our experimental setup 
in the UG laboratory, the coil has a diameter of about 10 cm,
about the same length, consists of several thousand turns
of insulated copper wire, and has a resistance of about 1000 $\Omega$.}
The positions of the weights $W_1$ and $W_2$ can be adjusted 
to bring the mean position of the bar magnet vertically below the
pivot O, and the position of coil is
adjusted so that its center coincides with this mean position 
($\theta=0$) of the magnet. The angular amplitude can be read by means of 
a scale and pointer. 
The magnet velocity can be controlled by choosing different angular amplitudes, 
\begin{figure}
\hspace*{11mm}
\psfig{file=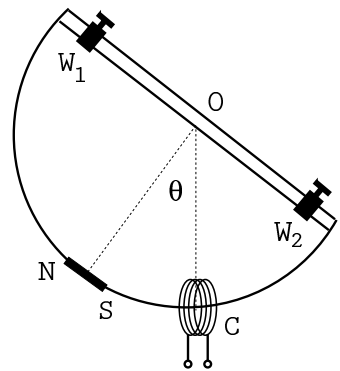,width=50mm,angle=0}
\caption{Experimental details.}
\end{figure}
\noindent
allowing the magnetic flux to be changed at different rates.

It is much more instructive to monitor the induced emf in the coil
through a PC interface,
which can be readily realized by low-cost, convenient data-acquisition 
modules available in the market. 
We have used a module based on the 
serial interface ``COBRA-3''  and its accompanying software
``Measure'', marketed and  manufactured by PHYWE.\cite{three}
We specially found useful the 
various features of ``Measure" 
such as ``integrate", ``slope", ``extrema", ``zoom" etc. 
In this article we describe modified experiments
designed to take advantage of the computer interface.
This allows for a quantitative and pedagogical study of
(i) angular (position) dependence of the magnetic flux 
through the coil, 
(ii) verification of Faraday's law of induction and 
(iii) electromagnetic damping, 
thereby revealing delicate features of the underlying mechanics.

\section{Induced emf pulse}
The equation for the induced emf ${\cal E}(t)$ as a function
of time $t$ can be written as
\begin{equation}
{\cal E}(t) = \frac{d\Phi}{dt}= 
\frac{d\Phi}{d\theta} 
\frac{d\theta}{dt}  \; ,
\end{equation}
expressing the combined dependence on the angular gradient $d\Phi/d\theta$  
and the angular velocity $\omega(\theta) = d\theta/dt$. 
This is reflected in the time dependence of ${\cal E}(t)$,
and a typical emf pulse is shown in Fig. 2;
the pulse-shape is explained below for one quarter
cycle of oscillation, starting from the extreme position
of the bar magnet ($\theta=\theta_0$).
As the magnet approaches the coil, the induced emf initially rises,
then turns over and starts falling as the magnet  
\begin{figure}
\vspace*{-70mm}
\hspace*{-28mm}
\psfig{file=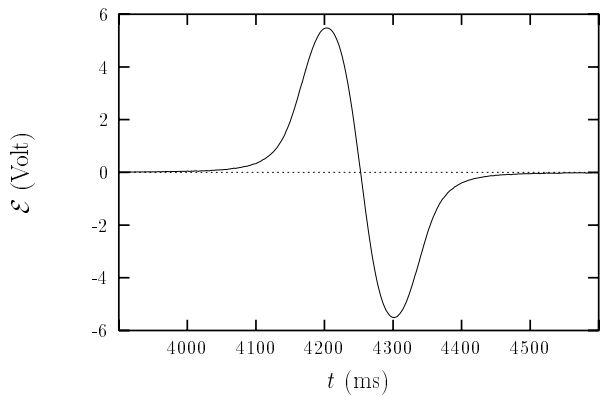,width=135mm,angle=0}
\vspace{-70mm}
\caption{A typical induced emf pulse.}
\end{figure}

\noindent
enters the coil and the magnetic flux begins to saturate, and finally changes sign
as the magnet crosses the center of the coil $(\theta=0)$ where
the flux undergoes a maximum.
Thus ${\cal E}$ actually vanishes
at the point where the angular velocity of the magnet is maximum. 

From Fig. 2 the magnitude of ${\cal E}$ is seen to be significant
only in a very narrow time-interval of about 200 ms,
which is much smaller than the oscillation time period
($T \approx 2$ s). 
This implies that the magnetic flux
through the coil falls off very rapidly
as the magnet moves away from its center,
so that $d\Phi/d\theta$ is significant only
in a very narrow angular range (typically about $5^\circ$)
on either side of the mean position.
As $d\Phi/d\theta = 0$ at $\theta=0$, it follows that
$d\Phi/d\theta$ is strongly peaked quite close to the
mean position, which accounts for the emf pulse shape 
seen in Fig. 2. 
This separation of the electromagnetic and mechanical time scales
has interesting consquences on the electromagnetic damping,
as further discussed in section IV.

\section{Magnetic flux through the coil}
In order to quantitatively study the magnetic flux through the coil,
the ``integrate'' feature of the software is especially convenient. 
From Eq. (1) the time-integral of the induced
emf directly yields the change in magnetic flux $\Delta \Phi$
corresponding to the limits of integration. If the lower limit 
of integration $t_i$ corresponds to the extreme position of the 
magnet ($\theta(t_i)=\theta_0$),\cite{id} where the magnetic flux through
the coil is negligible (valid only for large $\theta_0$), 
the magnetic flux $\Phi(\theta)$ for different angular positions 
$\theta(t)$ of the magnet is obtained as
\begin{equation}
\Phi(t) \approx
\int_{t_i} ^{t} {\cal E}(t') dt' \; .
\end{equation}

Figure 3 shows a plot of $\Phi(t)$ vs. $t$
for a large angular amplitude ($\theta_0 \sim 30^\circ $).
The time interval during which $\Phi(t)$ is significant
($\sim$ 200 ms) is a very small fraction of the oscillation time period
(about 2 sec), confirming that 
\begin{figure}
\vspace*{-70mm}
\hspace*{-36mm}
\psfig{file=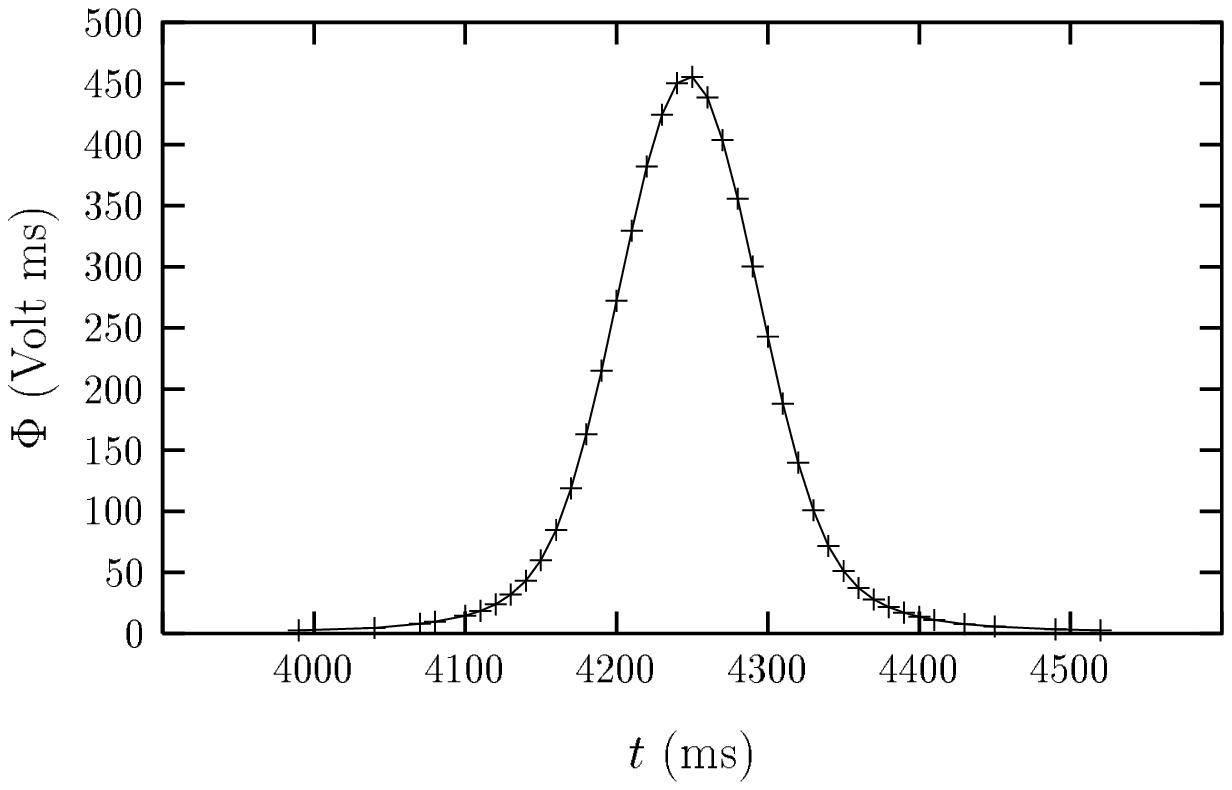,width=135mm,angle=0}
\vspace{-70mm}
\caption{Plot of the magnetic flux $\Phi$ through the coil with time $t$,
showing the rapid change as the magnet crosses the center of the coil.}
\end{figure}
\noindent
the magnetic flux changes very
rapidly as the magnet crosses the center of the coil.
As the angular velocity of the magnet is nearly
constant in the central region, the time scale in Fig. 3 can
be easily converted to the angular ($\theta= \omega t$) or linear 
($x=R\theta$) scale. The points of inflection, 
where $d\Phi/dt$ (and therefore $d\Phi/d\theta$) are extremum, 
are at 4200 and 4300 ms,
precisely where the peaks occur in the emf pulse (Fig. 2). 

\section{Verification of Faraday's law}
For  $\theta_0 >> 5^\circ$, 
the angular velocity of the bar magnet 
is very nearly constant in the narrow angular range 
near the mean position, and hence the 
peak emf ${\cal E}_{\rm max}$ is approximately given by
\begin{equation}
{\cal E}_{\rm max} \approx 
\left ( \frac{d\Phi}{d\theta} \right )_{\rm max}
\omega_{\rm max} \; .
\end{equation}
The maximum angular velocity $\omega_{\rm max}$ itself 
depends on $\theta_0$ through the simple relation (see Appendix)
\begin{equation}
\omega_{\rm max}=\frac{4\pi}{T} \sin (\theta_0/2) \; ,
\end{equation}
where $T$ is the time period of (small) oscillations.
Therefore if $\theta_0/2$ (in radians) is small compared to 
1, then $\omega_{\rm max}$ is 
nearly proportional to $\theta_0$, 
and hence ${\cal E}_{\rm max}$ approximately measures the 
angular amplitude $\theta_0$.

Eq. (3) provides a simple way for students to quantitatively verify 
Faraday's law. A plot of the peak emf ${\cal E}_{\rm max}$ 
(conveniently obtained using the software feature ``extrema") 
vs. $\omega_{\rm max}$ (evaluated from Eq. (4))
for different angular amplitudes 
should show a linear dependence (for large $\theta_0 $). 
While this behaviour could indeed be easily verified by students,
the interesting deviation from linearity expected
for low angular amplitudes $(\theta_0 \sim 5^\circ)$,
for which the $\theta$ dependence of the angular velocity 
$d\theta /dt$ is not negligible, 
turned out to be quite elusive. This 
\begin{figure}
\vspace*{-70mm}
\hspace*{-28mm}
\psfig{file=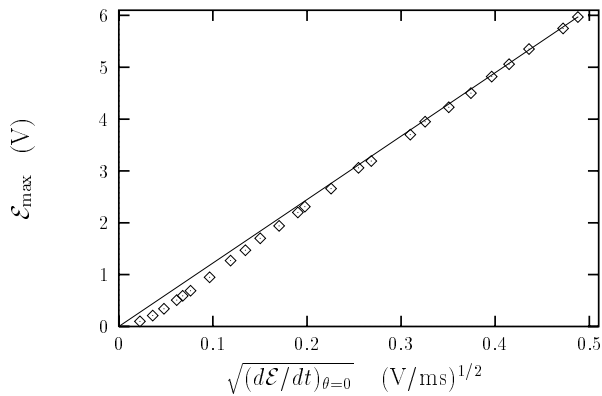,width=135mm,angle=0}
\vspace{-70mm}
\caption{
Plot of ${\cal E}_{\rm max}$ vs. 
$(d {\cal E}/dt)_{\theta =0} ^{1/2}$, 
showing the deviation from a straight line at low angular amplitudes.}
\end{figure}
\noindent
delicate deviation was presumably washed out by
the large percentage errors in $\theta_0$ measurements,
especially for small angles,  
precisely where this deviation is more pronounced. 

An alternative approach, which  
eliminates the need for measuring the oscillation amplitude 
$\theta_0$, is proposed below.
Taking the time derivative of the induced emf, 
and setting $\theta=0$, where the angular velocity is maximum,
we obtain 
\begin{equation}
\left ( \frac{d{\cal E}}{dt} \right )_{\theta = 0} =
\left ( \frac{d^2 \Phi}{d\theta ^2} \right )_{\theta=0}
\omega_{\rm max} ^2 \; ,
\end{equation}
which relates the slope of ${\cal E}$ 
at the mean position to $\omega_{\rm max} ^2 $.
As this relation holds for {\em all} amplitudes $\theta_0$,
it may be used to obtain $\omega_{\rm max}$ for
different angular amplitudes {\em without} the need for
any angular measurement. The slope at the mean position
(near zero crossing) is easily
measured through linear interpolation (see Fig. 2).
Thus, a plot of ${\cal E}_{\rm max} $ vs. 
$\sqrt{ (d{\cal E}/dt)_{\theta = 0} }$
should show both features of interest --- the  linear behaviour 
for large angular amplitudes and the deviation 
for very low amplitudes.
The key advantage of this plot lies in 
completely eliminating the errors associated with 
measurements of oscillation amplitudes $\theta_0$.

In this experiment the oscillations were started with a large initial
angular amplitude, and during the gradual decay of oscillations,
the peak voltages ${\cal E}_{\rm max} $
(on both sides of the mean position)
and the slope $ (d{\cal E}/dt)_{\theta = 0} $ were measured 
for a large number of pulses, so as to cover the full range 
from large to very small angular amplitudes. 
Fig. 4 shows this plot of the averaged
${\cal E}_{\rm max} $ vs. 
$\sqrt{ (d{\cal E}/dt)_{\theta = 0} }$ ,
clearly showing the deviation from linearity
for low angular amplitudes. To provide an idea of the
angular amplitude scale, the peak voltage
of $\sim 6 $ V corresponds to an angular amplitude 
$\theta_0 \approx 35^\circ$, so that the deviations 
become pronounced when $\theta_0 \approx 5^\circ$. 

\begin{figure}
\vspace*{-70mm}
\hspace*{-28mm}
\psfig{file=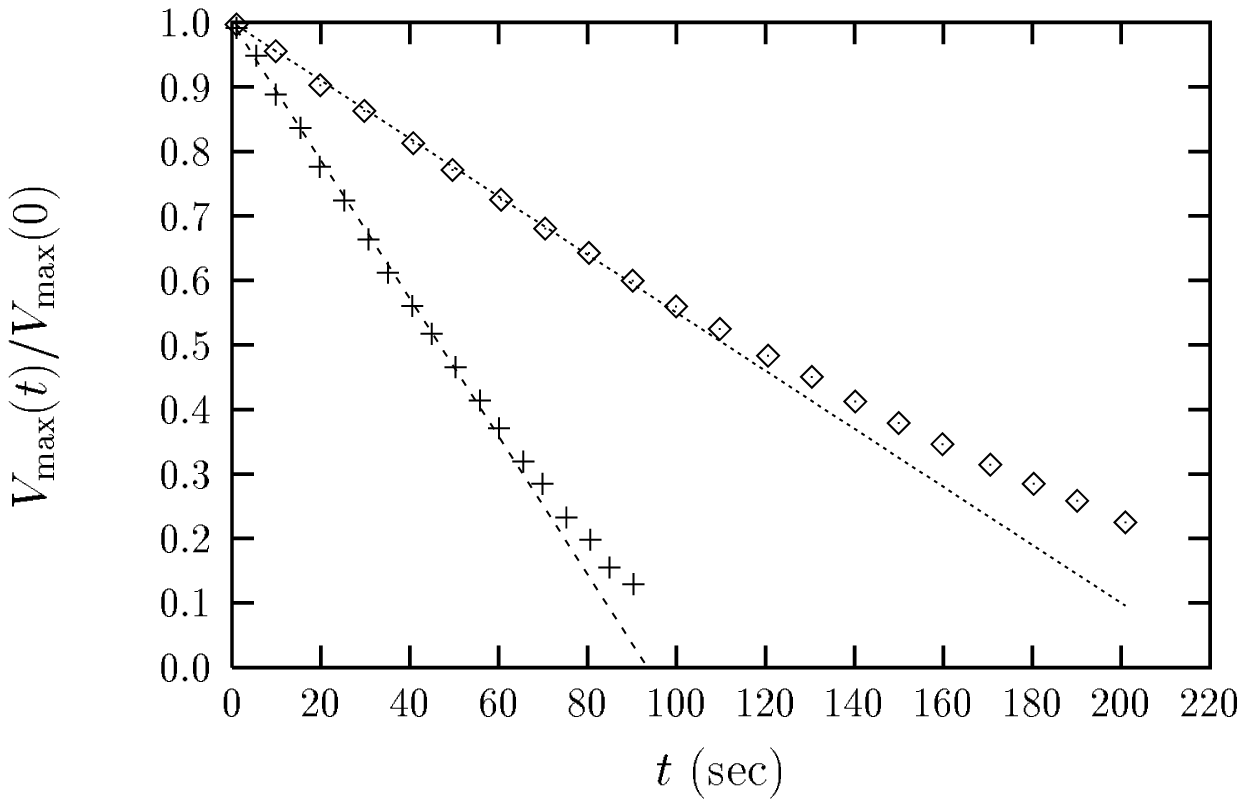,width=135mm,angle=0}
\vspace{-70mm}
\caption{
The normalized peak voltage $V_{\rm max}(t)/V_{\rm max}(0)$ vs. time $t$ 
for the short-circuit ($+$) and open-circuit ($\Diamond$) cases, 
showing nearly linear fall off.}
\end{figure}

\section{Electromagnetic damping}
To study the nature of the electromagnetic damping
in the oscillating system, the coil was short-circuited through 
a low resistance ($220\; \Omega$). The oscillations were started with
a large initial amplitude (still $\theta_0 /2 << 1)$, 
and the voltage $V(t)$ across 
the resistor was studied as a function of time.
As the oscillations decayed, the peak voltages $V_{\rm max}$
for sample pulses were recorded  at roughly equal intervals
on the time axis. This voltage
$V(t)$ is proportional to the current through the circuit, and hence
to the induced emf ${\cal E}(t)$. 
As the peak emf ${\cal E}_{\rm max}$ is approximately proportional
to the oscillation amplitude (except when the amplitude becomes 
too small), a plot of $V_{\rm max}$ vs. $t$ 
actually exhibits the decay of the oscillation amplitude with time.

Although an exponential decay of amplitude is more commonly
encountered in damped systems, the 
plot of the normalized peak voltage
$V_{\rm max}(t)/V_{\rm max}(0)$ vs. $t$ shows a distinctly 
linear decay (Fig. 5).
To distinguish the electromagnetic damping
from other sources (friction, air resistance etc.) 
the same experiment was repeated in the open-circuit configuration,
in which case electromagnetic damping is absent. 
In this case the amplitude decay is, as expected, much weaker, 
but significantly it is still approximately linear. 

A quantitative analysis of the energy loss
provides an explanation for this nearly linear decay in both cases.
We first consider the electromagnetic energy loss.
Neglecting radiation losses, the main source of energy loss 
is Joule heating in the coil due to the induced current.
Integrating over one cycle we have \\
\newpage
\begin{eqnarray}
\Delta E_{\rm one \; cycle} &=&
\int i^2 {\cal R} \; dt 
= \frac{1}{\cal R} \int {\cal E}^2 dt  \nonumber \\
&=&
\frac{1}{\cal R}
\int \left ( \frac{d\Phi}{d\theta}\right )^2
\left ( \frac{d\theta}{dt}\right )^2  dt \; ,
\end{eqnarray}
where $\cal R$ is the coil resistance.
This may be further simplified since 
$d\Phi/d\theta$ is significant only 
in a narrow angular range near 
$\theta= 0$ and rapidly vanishes outside.
Now, for amplitudes not too small, the angular velocity 
$d\theta/dt$ is nearly constant 
($\approx \omega_{\rm max}$) in this 
narrow angular range, and therefore 
taking it outside the integral, we obtain
\begin{equation}
\Delta E_{\rm one \; cycle} 
\approx  
\frac{\omega_{\rm max}}{\cal R} \int 
\left ( \frac{d\Phi}{d\theta}\right )^2 d\theta  \; .
\end{equation}
As the angular integral is nearly independent of 
the initial amplitude $\theta_0$, 
and therefore of $\omega_{\rm max}$,
the energy loss per cycle is proportional to 
$\omega_{\rm max}$, and therefore to  $\sqrt{E}$.
On a long time scale ($t >> T$),
we therefore have
\begin{equation}
\frac{d E}{dt} = -k \sqrt{E} \; .
\end{equation}

\noindent
Integrating this, with initial condition $E(0)=E_0$, we obtain
\begin{eqnarray}
\sqrt{E_0} - \sqrt{E}  \; & \propto &  \; \; t \nonumber \\
\Rightarrow \; \omega_{\rm max}^0 - \omega_{\rm max} 
\; \; & \propto & \; \; t \nonumber \\
\Rightarrow \; {\cal E}_{\rm max}^0 - {\cal E}_{\rm max} 
\; \; & \propto & \; \; t  \; ,
\end{eqnarray}
indicating linear decay of the peak emf,
and therefore of the amplitude, with time.

We now consider the energy loss in the open-circuit case,
where the damping is due to frictional losses.
A frictional force proportional to velocity,
as due to air resistance at low velocities, will result 
in an exponential decay of the oscillation amplitude.
However, a function of the type $e^{-\alpha t}$ did not
provide a good fit.
On the other hand, assuming a constant frictional torque
$\tau$ at the pivot, which is not unreasonable 
considering the contact forces at the pivot, 
we obtain for the energy loss in one cycle,
\begin{eqnarray}
\Delta E_{\rm one \; cycle} &=&
\int \tau \;  d\theta  = \tau \; 4\theta_0 \nonumber \\
&\propto & \; \omega_{\rm max} \nonumber \\
&\propto & \;  \sqrt{E} \; .
\end{eqnarray}
This is similar to the earlier result of Eq. (8) for electromagnetic damping,
yielding a linear decay of the oscillation amplitude with time,
which provides a much better fit with the observed data,
as seen in Fig. 5. 
The deviation from linearity at large times is presumably due to a
small air resistance term. 
In fact, if a damping term $-k'E$ due to air resistance is included in Eq. (8),
the differential equation is easily solved, 
and the solution provides an excellent fit to the data.
Finally, another term $-k'' E^{3/2}$ should be included in Eq. (8), arising from 
the reduction in the average centripetal force $Ml\omega^2 \theta_0 ^2 /2$ 
with the oscillation amplitude $\theta_0$,
which decreases the frictional force at the pivot
due to the reduction in the normal reaction.

\section{Summary}
A pedagogically instructive study of electromagnetic induction and damping 
is made possible by attaching a PC interface 
to a conventional setup for studying Faraday's law.
By eliminating all errors associated with angular measurements,
the novel method applied for the verification of Faraday's law
reveals delicate features associated with the underlying mechanics.
A quantitative analysis of the distinctly linear decay of 
oscillation amplitude due to electromagnetic damping 
reveals an interesting interplay of the
electromagnetic and mechanical time scales. 

\section{Appendix}
If the system is released from rest with an angular displacement $\theta_0$,
then from energy conservation 
\begin{equation}
\frac{1}{2}I \omega_{\rm max} ^2 =
Mgl(1-\cos \theta_0) \; ,
\end{equation}
where $M$ is the mass of the system,
$I$ its moment of inertia about the pivot O, and $l$ the distance
from O to the centre of gravity.
For small oscillations, the equation of motion is
$I\ddot{\theta} = - (Mgl) \theta $, so that the time period is given by
\begin{equation}
T = 2\pi \sqrt{ \frac{I}{Mgl } } \; .
\end{equation}
Eliminating $I/Mgl $ from these two equations, we obtain
\begin{equation}
\omega_{\rm max}  =
\frac{4\pi}{T} \sin (\theta_0/2) \; .
\end{equation}

\section*{Acknowledgement}
Helpful assistance from Shri B. D. Gupta, B. D. Sharma, and Manoj Kumar
is gratefully acknowledged.


\begin{references}
\bibitem[6]{id}
Identifying the time $t_i$
for a given induced emf pulse ${\cal E}(t)$
is a good exercise for students.
\end{references}
\end{document}